# Efficiency of Using Utility for Usernames Verification in Online Community Management

Solomiia Fedushko*[0000-0001-7548-5856], Yuriy Syerov[0000-0002-5293-4791],
Oleksandr Skybinskyi [0000-0002-8459-1420], Nataliya Shakhovska[0000-0002-6875-8534],
Zoryana Kunch [0000-0002-8924-7274]

Lviv Polytechnic National University, S. Bandera str. 12, 79013 Lviv, Ukraine

`solomiia.s.fedushko@lpnu.ua, yurii.o.sierov@lpnu.ua, nataliya.b.shakhovska@lpnu.ua`

**Abstract.** The study deals with the methods and means of checking the reliability of usernames of online communities on the basis of computer-linguistic analysis of the results of their communicative interaction. The methodological basis of the study is a combination of general scientific methods and special approaches to the study of the data verification of online communities in the Ukrainian segment of the global information environment. The algorithm of functioning of the utility "Verifier of online community username" is developed. The informational model of the automated means of checking the usernames of online community is designed. The utility "Verifier of online community username" data validation system approbation is realized in the online community. The indicator of the data verification system effectiveness is determined.

**Keywords:** online community, username, community management, content, verification, information, efficiency.

## 1 Introduction

Taking into account the trends and dynamics of the development of online communities in the Ukrainian segment of the global information environment, the development of software tools for managing the web communities is a priority. Despite the fact that online communities accumulate a large amount of data and become more popular among web users, web verification software is still incomplete, formal, superficial, and reduces to data matching and checking for certain formal features (for example, software tools that are designed for video verification, verification of veracity of metadata and search for users in social networks, etc.). The scientific task of developing methods and means of verifying the authenticity of personal data of users of online communities, in particular their nicknames of online users, on the basis of computer-linguistic analysis of information content is an actual direction of scientific research in the field of computer linguistics.

The program implementation of verifying the authenticity of the data of online communications' users by means of computer-linguistic analysis of information traces of online users is one of the important tasks of researches on the Internet, mathematical linguistics and related scientific fields.

*The purpose of the study.* The purpose of the work is to develop new methods and means of checking the reliability of the usernames of online communities by the results of computer-linguistic analysis of information traces of users in online communities.

*The object of research.* The object of research is the processes of communicative interaction of users of online communities.

*The subject of the study.* The subject of the study is the methods and means of checking the reliability of usernames of online communities on the basis of computer-linguistic analysis of the results of their communicative interaction.

*Research methods.* Studies performed during work on the basis of applied methods and mathematical linguistics (structural analysis methods for studying the information content of users in online communities and methods of analyzing online content aimed at the study of verbal characteristics of texts). Simulation of the information scheme of the dictionary and software for verifying the nicknames of the user of the online community is accomplished with the help of diagrammatic structural modeling tools.

## 2  Development of utility for verifying online community username

Developing a new approach to username validation, provided by the user of online community during registration, is an urgent issue in managing e-government, distance learning and moderating the online community, media resources, social networks, encyclopedias, blogs, and more. According to the increasing number of users of these resources in the network and the popularity of social communications there is a need to develop a method of computer-linguistic verification of personal data by social communications. One of the areas of research is to develop methods for validating the username of a potential participant in a web service.

To perform the computer-linguistic verification of the usernames of the online community, the algorithm of functioning of the utility "Verifier of online community username" was developed (see Fig. 1).

The algorithm of this utility is intended for verification of registration data in online communities and has been successfully implemented in the work of several online communities.

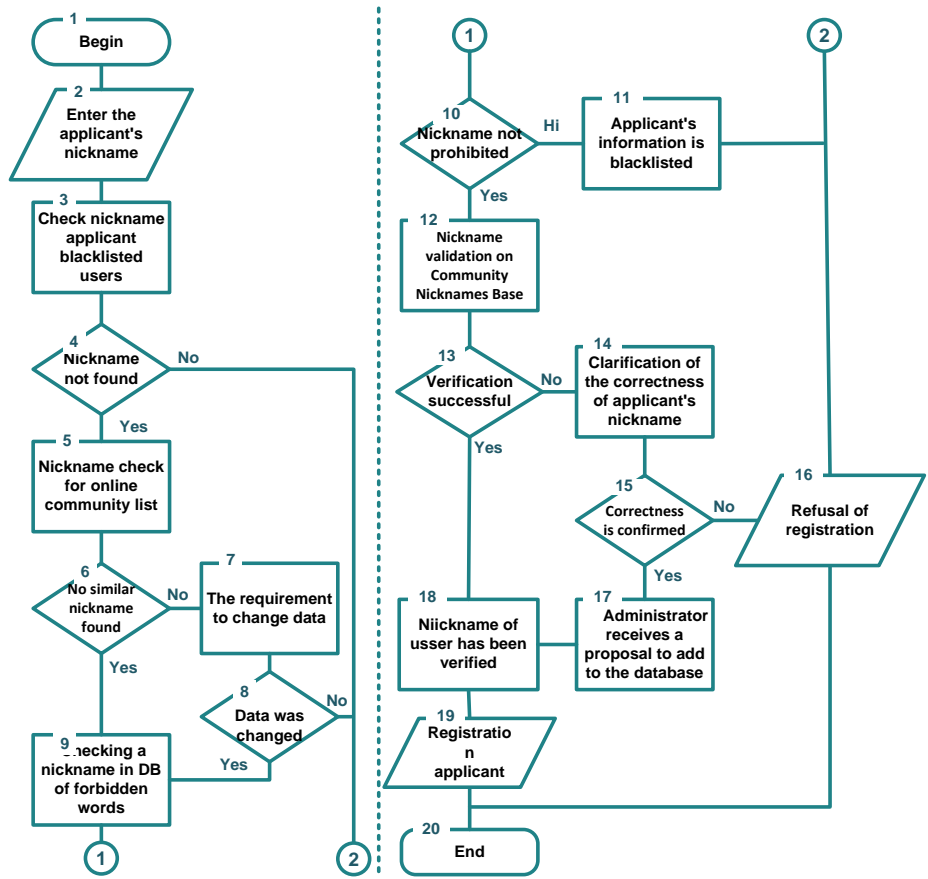

**Fig. 1.** The scheme of algorithm of utility "Verifier of online community username" process

— Checking the username field in the online community registration form.

The online community registration form is required to specify a username. Username is the unique name of the web member in the online community. The choice of username depends on the level of communicative behavior of the user in the online community, the authority of the user among other users of the online community.
Optimal for the effective functioning of the online community are the requirements for the formulation of the name of members of the online community is format as "First Name. Last Name". The process of checking the online community username field for compliance with the "First Name.Last Name" is includes checking the "username" field according to the algorithm of processing utility "Verifier of online community username".

— Checking informationn in the online community database.

According to the algorithm of process of utility "Verifier of online community username" the Internet username verification is performed in the following databases:

- *DB "Prohibited content"* is created by administrators at the beginning of the creation of the online community according to the established rules in the online community.
- *DB "Blacklist of online community users"* is a list of all names of community users, who have chosen an internet name that is inappropriate with the online community's rules of procedure and is not desired to change the data to correct in the stage of validation of Internet user name. This list is designed to save time of administrators and moderators, to reduce the cost of community moderation.
- *DB "Usernames list of online community users"* is a list of names of all registered online community users that automatically determines the usernames' correctness, defines the spelling language, and avoids duplication of the names of online community members, which will not cause problems with the identification of online community members. Administrators regularly update these databases.

– Results of the analysis of the username of the online community.

If the user followed all the rules of the community and provided accurate information, it successfully completed the registration. Otherwise, the user will hardly be denied registration.

Functioning utility "Verifier of online community username" solves the problem of automating the process of validating usernames of online community. Information automated model of utility for checking online community usernames "Verifier of online community username" described in further work.

Specialized utility "Verifier of online community username" is designed for preregistration and post-registration verification of Internet names of users of the online community. The name of the online community users indicates when registering in the community and automatically places an account and is covered under each post by the author in communicating with other users of the online community.

A method for verifying the Internet name of a user in the online community has been developed as the basis for the implementation of the utility "Verifier of online community username".

The software for the analysis of Internet names solves the following tasks:

- online registration of the online name in order to avoid duplication of names;
- filtering registered web attendees according to the criteria for choosing Internet names in accordance with the rules introduced in the online community;
- verification of the correctness of the identification of such personal data as the name and surname of the user, the geographical location, the e-mail address and all additional contact information;
- post-registration computer-linguistic analysis of avatar, browser, biography, signature and web user status.

So, an important mechanism for verifying the veracity of web users' data is the use of an automated checking tool for the online user of the online community. We model the information model of this automated means with the help of the unified data representation tool - the diagrams (see Figure 2).

The information model is executed according to Barker's notation and contains 5 entities that are linked by "one-to-many" links. The "Blacklist of Users" essence contains information about users of the online community that violated the system of rules of communicative behavior of users of the online communities and to which the community administration has applied the highest level sanctions. The most important attributes: the "User's Online Name" attribute contains the Internet name of the users of the Sun and the attribute "Sanctions Code" contains information on the sanctions that are introduced to the Sun users.

The "User" essence contains the basic information about the user of the online community: the class and level of the anonymity of the user, the internet-name code, contacts and the date of registration.

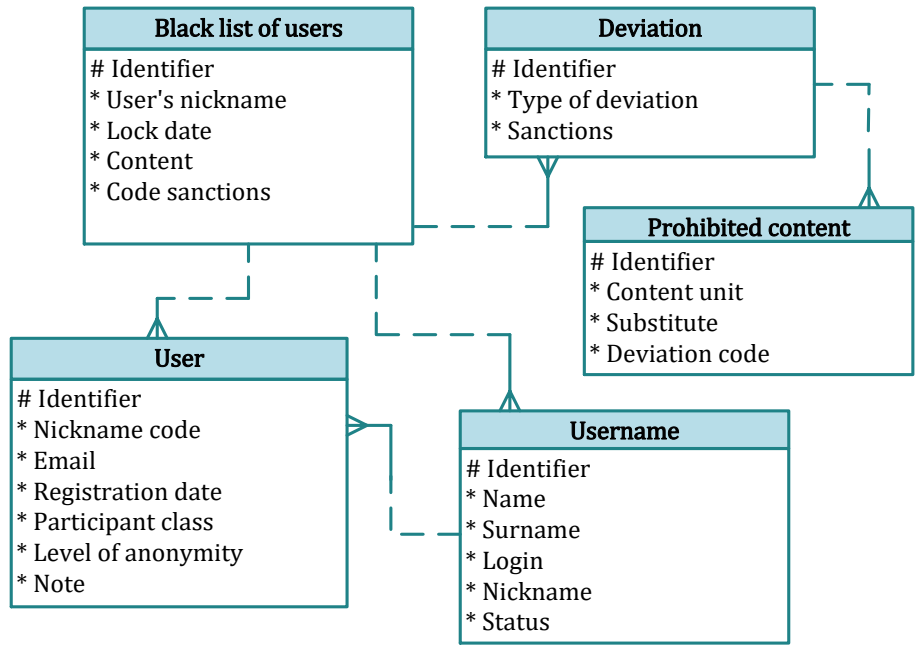

**Fig. 2.** Informational model of the automated means of checking the usernames of online community

The essence of Deviation includes information on violations of the rules of communicative behavior of users of online communities and the application of appropriate sanctions contained in the Sanctions Attribute.

The "Prohibited Content" essence contains information about all possible prohibited content for publishing in the online community.

All information about the names of the users of the online community is contained in the "Username" entity.

According to the information model, the effectiveness of the results of the automated means of checking the Internet name of the user utility "Verifier of online community username" directly proportional to the filling of the three databases. The filling of these databases is the result of the computer-linguistic validation of Internet user names of online communities.

The functionality of utility "Verifier of online community username" is based on the following components:

- Component of registration and validation of personal data;
- Component of the analysis of usernames of online community.

### 2.1 Registration and validation component of personal data

Registration and validation component of personal data is intended for validation of the data both at the registration of the user, and already registered user, who for some time has already participated in network communication with other online community users. This verification of the truthfulness and correctness of data is realized in accordance with the algorithm of registration and validation.
The component performs the following tasks:

- classification of online users by the authenticity of personal data;
- introduction of the new registration methods, which allowed to eliminate the soon-to-be problematic users of the community;
- development of a validation method of personal data, which allows you to select users whose actions need to be permanently monitored.

The automated solution to these tasks facilitates more efficient functioning of the online communities and systematization of their moderation (administration) process, which, as a result, reduces the time and money, spent on management and increases the competitiveness of the online community.

### 2.2 Analysis component of the online users' Internet names

Component's functioning is based on the algorithm of the computer-linguistic validation of the online community users' Internet names. The basic analysis stages of the online community users' Internet names are demonstrated in the component's functioning scheme of utility "Verifier of online community username" (see Fig. 3).

The primary task of the utility "Verifier of online community username" is to check the data availability in such databases: forbidden content, blacklist of the VC users and the online community usernames list.

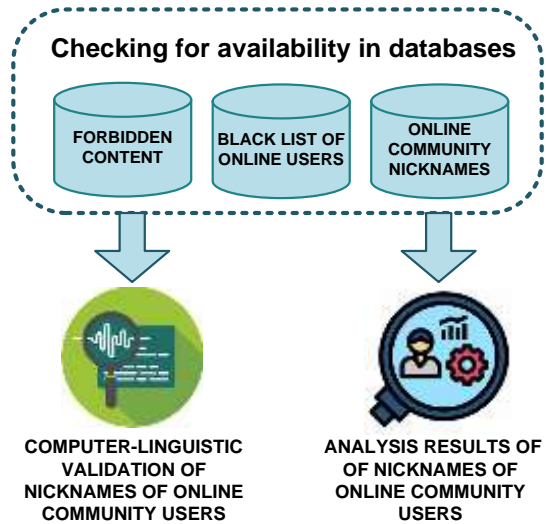

**Fig. 3.** The functioning scheme of utility "Verifier of online community username"

The user interface of the utility "Verifier of online community username" is shown in Fig. 4.

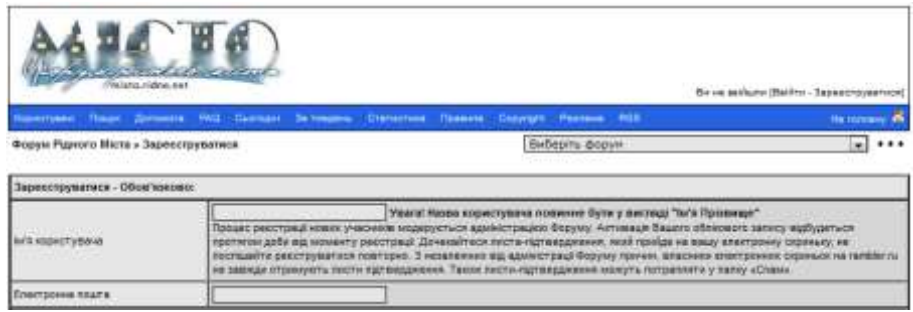

**Fig. 4.** The utility "Verifier of online community username" interface

The result of a result of registration process of usernames is shown in Fig. 5.

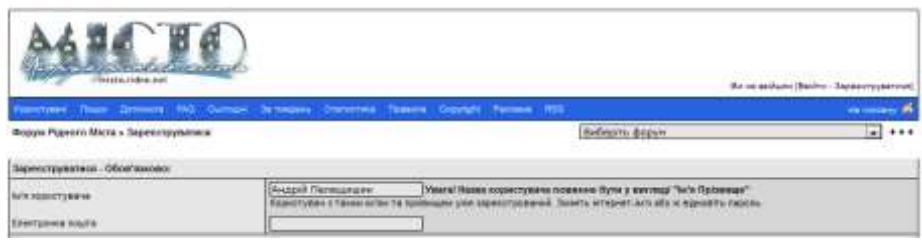

**Fig. 5.** The result of registration process of usernames of online community

This component's results are the outputs of the utility "Verifier of online community username" and the report based on them is generated for the moderator (the online community administrator).

Based on this report, when data verification level is low or the expectancy of false personal information in the account is high, or when Internet name is incorrect, the administrator may choose to automatically send a warning message with the requirement to provide reliable data, or even block the user account of the online community.

## 3    Research results approbation

The priorities of the online community's functioning are determined by the owners at the project development stage, which depends on the chosen topic and the community development scenario. The task of filtering the users of online community on the basis of truthfulness and data correctness emphasizes the quality and validity of the online community's information content, and the number of users is already a secondary factor. The users' personal data verification resulted in the transition of online communities to a qualitatively new functioning stage, where the priority is used by active online users who provide personal data with a high level of reliability and authenticity in general and the Internet name in particular.

Use the results to improve popular online communities' management. One of the important results of this work is the development and program realization of the software complex of computer-linguistic reliability analysis of the online user's socio-demographic characteristics. The utility "Verifier of online community username" data validation system approbation was realized in the online community.

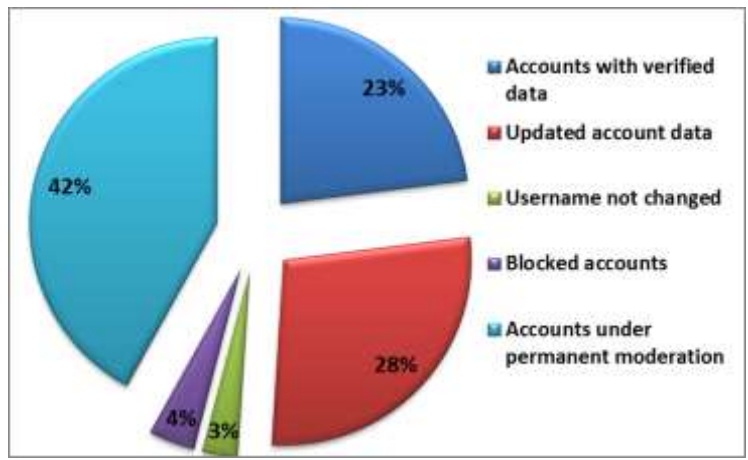

**Fig. 6.** Online community users' classification

The results show that 23% of the online community users (out of a total of 4708 users) provided reliable information in their accounts. 28% of users updated their credentials in the accounts, although 3% did not change their internet names at the ad-ministrators' request. However, due to their authority in the community and active and high-quality Internet communication, the administration did not apply sanctions to them. 4% of community users' accounts are blocked due to online community rules violation and actions that affect the effective functioning of the online community. 42% of the accounts are under constant moderation.

The effectiveness of the verification system shows that the workload for the moderators of the personal data verification is reduced by 2-3 times depending on the online community type, respectively, and the reduction of time and financial costs for administering online communities is essential.

The indicator of the data verification system effectiveness is determined as follows:

$$\text{Efficiency} = \frac{N^{VerPD}}{N^{VerPD} - N(LAdequacy)^{APD}}, \quad N^{VerPD} \neq N(LAdequacy)^{APD}, \quad (1)$$

where $N(LAdequacy)^{APD}$ is the number of online based user accounts with low data adequacy, $N^{VerPD}$ is the total number of verified online users' accounts.

Compared with the practice of expert verification of the online communities' users' personal data authenticity, the results can increase the effectiveness of the overall process of their management by 20-30% depending on the online community specifics.

## 4 Conclusions

There was created the complex of means of authenticity checking of usernames of online community by verifying the users' data and theirs information content of web.

The task of the utility is to increase the moderation efficiency of online communities in the following directions: online community management, cybersecurity, targeting of internet-advertising, profitability of the community support project. Thus, consumers of the utility "Verifier of online community username" can be owners of online communities. The utility involves execution of individual tasks in an automated mode, but in any case, the results of its activities are informative.

Basically, the utility is a decision support mean for a community management specialist in the field of processing the personal data of users in the online community.

Requirements to the software complex are formulated on the basis of the international standard ISO-9126, which is acknowledged in the world and in Ukraine and concerns the quality and efficiency of software products. In particular, the standard defines the following criteria for software systems: functional possibilities, reliability, practicability, efficiency, extension, and mobility. Some features of the means are solved within the existing today software packages or libraries (for example, basic parsing of hypertext – parsing). Implementation or algorithmic description of some applied elements of the complex, which go beyond the scope of research, is available only for individual human languages (most often English). However, provided that such implementations are introduced for other languages, they can be implemented in the complex.